\documentclass[conference]{IEEEtran}
\IEEEoverridecommandlockouts
\usepackage{cite}
\usepackage{algorithmic}
\usepackage{textcomp}
\usepackage{xcolor}
\usepackage{graphicx}
\usepackage{multirow}
\usepackage{url}
\usepackage{comment}
\usepackage{todonotes}
\usepackage{amsmath, amsthm, amssymb, latexsym, amsfonts}
\usepackage{multicol}
\usepackage{subfig}
\usepackage{booktabs}
\usepackage{csquotes}
\usepackage{float}
\usepackage[hidelinks,colorlinks=true,linkcolor=blue,citecolor=blue]{hyperref}

\def\BibTeX{{\rm B\kern-.05em{\sc i\kern-.025em b}\kern-.08em
    T\kern-.1667em\lower.7ex\hbox{E}\kern-.125emX}}

\begin{document}

\title{Assessing Encoder-Decoder Architectures for Robust Coronary Artery Segmentation}

\author{%
  Shisheng Zhang\IEEEauthorrefmark{1},
  Ramtin Gharleghi\IEEEauthorrefmark{1},
  Sonit Singh\IEEEauthorrefmark{2},
  Arcot Sowmya\IEEEauthorrefmark{2},
  Susann Beier\IEEEauthorrefmark{1}
  \\
  \IEEEauthorblockA{%
    \IEEEauthorrefmark{1} School of Mechanical and Manufacturing Engineering, University of New South Wales, Sydney, Australia
  }
   \IEEEauthorblockA{%
    \IEEEauthorrefmark{2} School of Computer Science and Engineering, University of New South Wales, Sydney, Australia
  }
 
  Email: \url{shisheng.zhang@unsw.edu.au} 
}

\IEEEoverridecommandlockouts
\IEEEpubid{\makebox[\columnwidth]{~\copyright2023 IEEE. Article accepted for publication in IVCNZ 2023.} \hspace{\columnsep}\makebox[\columnwidth]{}}

\maketitle
%\IEEEpubidadjcol

\begin{abstract}
Coronary artery diseases are among the leading causes of mortality worldwide. Timely and accurate diagnosis, facilitated by precise coronary artery segmentation, is pivotal in changing patient outcomes. In the realm of biomedical imaging, convolutional neural networks, especially the U-Net architecture, have revolutionised segmentation processes. However, one of the primary challenges remains the lack of benchmarking datasets specific to coronary arteries. However through the use of the recently published public dataset ASOCA, the potential of deep learning for accurate coronary segmentation can be improved. This paper delves deep into examining the performance of 25 distinct encoder-decoder combinations. Through analysis of the 40 cases provided to ASOCA participants, it is revealed that the EfficientNet-LinkNet combination, serving as encoder and decoder, stands out. It achieves a Dice coefficient of $0.882$ and a $95$th percentile Hausdorff distance of $4.753$. These findings not only underscore the superiority of our model in comparison to those presented at the MICCAI 2020 challenge but also set the stage for future advancements in coronary artery segmentation, opening doors to enhanced diagnostic and treatment strategies.
\end{abstract}


%%%%%%%%%%%%%%%%%%%%%%%%%%%%%%%%%%%%%%%%%%%
%%%%%%%%%%%%%%%%%%%%%%%%%%%%%%%%%%%%%%%%%%%

\section{Introduction}~\label{Introduction}
Coronary artery disease is one of the leading causes of mortality globally~\cite{Roth:2017:global_regional_national_burden}, and has been recognised as a major health concern by the United Nations~\cite{who}. Marked by the narrowing of the coronary luminal area, this condition restricts the blood supply to the heart muscle, which can lead to ischemia, heart attacks and death~\cite{Wackers:1976:CAD,Libby:2005:Pathophysiology_CAD}. 

Computed Tomography Coronary Angiography (CTCA) provides insights into artery conditions. Patients undergoing CTCA are administered a contrast agent to increase the contrast between blood vessels and the background in images. This enhanced distinction facilitates the accurate identification of obstructions or the assessment of arterial health. Within this context, automated segmentation emerges as an essential tool. The incorporation of automated segmentation not only augments clinical diagnostic accuracy but also drives advancements in the realm of cardiovascular imaging research~\cite{Djukic:2013:virtual_reality,Gharleghi:2022:automated_segmentation}. Given the extensive range of potential applications for automated segmentation, several methods have been proposed~\cite{Gharleghi:2022:Towards_automated_CAD}. However, most of these methods have been developed using private datasets, with only a few tested on the Rotterdam dataset, which is currently unavailable~\cite{Schaap:2009:standardized_evaluation_methodology}. This lack of access makes it difficult to benchmark different models effectively. The recently available ASOCA dataset is a notable exception, which has not been tested by many automated segmentation works~\cite{Gharleghi:2023:Annotated_CTCA}. The ASOCA public dataset has 30 normal and 30 disease cases with annotations from CTCA scans. It includes anonymised CTCA images combined with high-quality manual voxel annotations derived from 3 experts. Other coronary arteries datasets are often limited by a lack of annotations and specific usage restrictions. However, ASOCA dataset is freely available to researchers, has high-quality annotations, and has a balanced distribution of normal and disease cases. 

A challenge based on the ASOCA dataset organised in MICCAI 2020 showed that algorithms based on U-Net performed quite effectively for coronary artery segmentation~\cite{Gharleghi:2022:automated_segmentation}. However, many renowned convolutional neural networks, particularly those functioning as 3D encoders and decoders, have not been extensively tested on ASOCA. Therefore, this work will benchmark the dataset using a few combinations of convolutional encoders and decoders. This work fills the gap of exploring and assessing new segmentation architectures for the first open-source dataset in coronary artery segmentation and advocates for benchmarking with the ASOCA dataset in the research community. 

%%%%%%%%%%%%%%%%%%%%%%%%%%%%%%%%%%%%%%%%%%%
%%%%%%%%%%%%%%%%%%%%%%%%%%%%%%%%%%%%%%%%%%%
\section{Related Work}\label{related_Work}
Most recent methods leverage advances in deep learning and computational power to segment arteries, particularly through the use of convolutional neural networks (CNNs). CNNs involve training a model with convolutional blocks using gradient descent to predict whether a voxel is part of the coronary artery tree. The majority of CNNs for biomedical image segmentation draw inspiration from the U-Net architecture, which consists of convolutional layers applied to progressively down-sampled versions of the image~\cite{Ronneberger:2015:U-Net}. This allows the model to gather features at multiple scales. These features are then up-sampled and combined with higher-resolution images to make the final prediction. 

Starting from the basic U-Net structure, more advanced CNNs have been employed for coronary artery segmentation. For instance, dual-CNN models have been used. The first CNNs identify voxels, and the second CNNs segment the artery from the output of the first~\cite{Chen:2018:automatic_coronary_artery}. Another study attempted to use a CNN in conjunction with a Recurrent Neural Network (RNN) to identify potential arteries~\cite{Mirunalini:2019:segmentation_of_CA}. To compensate for the lack of medical data, weights pre-trained on Resnet-50 were used in a V-Net model for artery segmentation~\cite{Fu:2020:MaskR-CNN_based_CAD}. Moreover, the attention gate has also been applied in deep learning models~\cite{Shen:2019:CA_segmentation,Lei:2020:automated_CA_segmenation}. The centerlines of arteries, a crucial component of coronary arteries, have been either used as supplementary data for model training or designed as output targets to guide the training process~\cite{Hong:2017:coronary_luminal_and_wall_mask,Huang:2020:coronary_wall_segmentation,Kong:2020:learning_tree_structured_representation}. Various post-processing methods have been attempted, including active contour for model refinement, and region growing to finalise the segmentation~\cite{Gu:2020:segmenation_of_coronary_arteries,Tian:2021:automatic_coronary_artery_segmentation}. 

Reviewing recent work reveals that despite the combination of CNNs and RNNs and modifications of pre- or post-processing methods, not much attention has been paid to complex encoders or decoders to enhance the accuracy of artery segmentation. In this work, we aim to assess a number of recent deep learning-based 3D encoders and decoders used to extract coronary artery segmentation from CTCA. 

%%%%%%%%%%%%%%%%%%%%%%%%%%%%%%%%%%%%%%%%%%%
%%%%%%%%%%%%%%%%%%%%%%%%%%%%%%%%%%%%%%%%%%%
\section{Methodology}\label{methodology}

\subsection{Dataset}
The dataset comprises 30 healthy individuals without stenosis and non-obstructive disease, and 30 patients with obstructive disease and evidence of calcium scores greater than 0~\cite{Gharleghi:2023:Annotated_CTCA}. The images were created utilising retrospective electrocardiogram gating on a multi-detector CT scanner (GE Lightspeed 64 multi-slice scanner, USA). Beta-blockers were administered in order to keep the patient’s pulse rate at or below 60 beats per minute while imaging with a contrast medium (Omnipaque 350). The images were exported as DICOM files, and the end-diastolic time step was retained for analysis. The resolution on the z-axis is 0.625 mm, while the in-plane resolution varies between 0.3mm and 0.4 mm based on the patient. More details on this preceding work can be found in Gharleghi et al.~\cite{Gharleghi:2023:Annotated_CTCA}. 

\begin{figure*}
    \centering
    \includegraphics[scale=0.67]{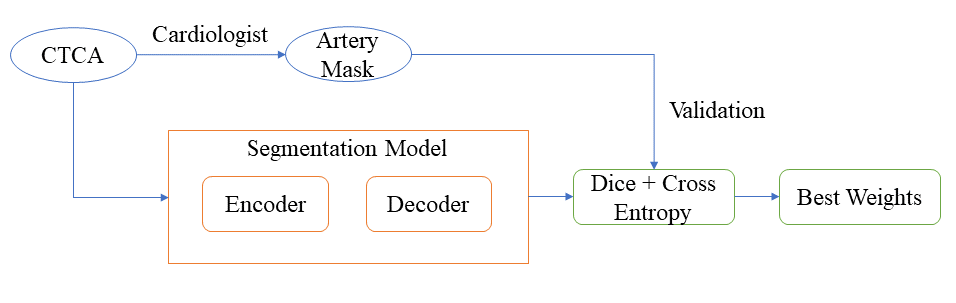}
    \caption{Flowchart of assessment method.}
    \label{fig:flowchart}
\end{figure*}

\subsection{Deep learning model development}
An illustration of a complete assessment system is shown in Figure~\ref{fig:flowchart}. We used the segmentation model consisting of an encoder and decoder to distinguish coronary arteries on the medical images. Five-fold cross-validation was used to validate the models. Best weights are saved with the highest Dice coefficient. We used encoders architectures from  EfficientNet~\cite{Tan:2019:EfficientNet}, ResNet~\cite{He:2016:ResNet}, InceptionNet~\cite{Szegedy:2016:Inception}, and DenseNet~\cite{Huang:2017:DenseNet} to generate feature maps, with the structures shown in Table~\ref{tab:model_structures}. Patches of dimensions $64 \times 64 \times 64$ are fed into the input layer. These images then traverse through four layers of the designed architecture, extracting the most significant features. Each encoder layer maintains a distinct feature map dimension.  In our case, an input size of $64 \times 64 \times 64$ can offer a balance between computational efficiency and the required resolution to detect relevant features. It ensures consistency across variable-size scans and enables the model to consider a 3D spatial context, which is essential for accurate predictions. 

\begin{table*}[]
    \centering
    \caption{Model structures of convolutional encoders. Note that input layer is $1 \times 64 \times 64 \times 64$}
    \label{tab:model_structures}
    \begin{tabular}{lccccc}
    \toprule
    \textbf{} & \textbf{EfficientNet} & \textbf{ResNet} & \textbf{Inception} & \textbf{DenseNet} & \textbf{U-Net Encoder} \\
    \midrule
    Layer 1 & $128 \times 32 \times 32 \times 32$ & $256 \times 32 \times 32 \times 32$ & $32 \times 32 \times 32 \times 32$ & $64 \times 32 \times 32 \times 32$ & $64 \times 32 \times 32 \times 32$ \\
    Layer 2 & $256 \times 16 \times 16 \times 16$ & $512 \times 16 \times 16 \times 16$ & $128 \times 16 \times 16 \times 16$ & $128 \times 16 \times 16 \times 16$ & $128 \times 16 \times 16 \times 16$ \\
    Layer 3 & $512 \times 8 \times 8 \times 8$ & $1024 \times 8 \times 8 \times 8$ & $512 \times 8 \times 8 \times 8$ & $192 \times 8 \times 8 \times 8$ & $256 \times 8 \times 8 \times 8$ \\
    Feature Map & $1024 \times 4 \times 4 \times 4$ & $2048 \times 4 \times 4 \times 4$ & $512 \times 4 \times 4 \times 4$ & $256 \times 4 \times 4 \times 4$ & $256 \times 4 \times 4 \times 4$ \\
    \bottomrule
    \end{tabular}
\end{table*}

Afterwards, the DeepLabV3~\cite{chen:2017:rethinking}, LinkNet~\cite{Chaurasia:2017:LinkNet}, Feature Pyramid Network~\cite{Lin:2017:feature_pyramid_networks}, and Pyramid Attention Network~\cite{li:2018:pyramid} were further used to process the feature maps generated by the encoders. The decoder’s input layers are adjusted according to the size of the feature maps with the structures shown in Table~\ref{tab:architectures}. Finally, the decoders generate a mask distinguishing arteries from non-artery regions, represented as a $64 \times 64 \times 64$ array that shows a 3D segmentation mask. The combination of both decoder and encoder architectures is shown in Table~\ref{tab:architectures}. The model’s computations are carried out using FP32 precision. 

\begin{figure*}
    \centering
    \includegraphics[scale=0.7]{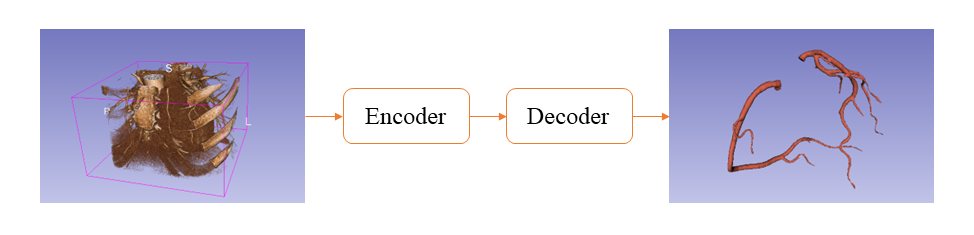}
    \caption{Diagram of segmentation model.}
    \label{fig:segmentation_model}
\end{figure*}

Twenty-five configurations for experiments were examined, and an example of the raw input and output of the segmentation model is shown in Figure~\ref{fig:segmentation_model}. The advancements in segmentation tasks have recently been attributed more to the encoder-decoder architecture, as evidenced by the work on U-Net, 3D U-Net and their variants~\cite{Ronneberger:2015:U-Net,Cicek:2016:3D-UNet}. This structure was influenced by the Convolutional Neural Network, introduced in 1998, but went further to incorporate a decoder network, providing a robust solution for pixel-wise prediction~\cite{Lecun:1998:gradient_based_learning}. 

\begin{table*}
\centering
\caption{The architecture of segmentation models}
\label{tab:architectures}
    \begin{tabular}{llrr}
    \toprule
    \textbf{Encoder} & \textbf{Decoder} & \textbf{\# of trainable parameters} & \textbf{Model size (MB)} \\
    \midrule
    \multicolumn{1}{l}{EfficientNet} & \multirow{4}{*}{DeepLabV3} & \multicolumn{1}{r}{12,275,848} & \multicolumn{1}{r}{722.31} \\
    \multicolumn{1}{l}{ResNet} &                & \multicolumn{1}{r}{51,372,456} & \multicolumn{1}{r}{2547.14} \\
    \multicolumn{1}{l}{InceptionNet} &          & \multicolumn{1}{r}{14,597,944} & \multicolumn{1}{r}{429.20} \\
    \multicolumn{1}{l}{DenseNet} &              & \multicolumn{1}{r}{5,780,928} & \multicolumn{1}{r}{1243.41} \\
    \multicolumn{1}{l}{U-Net Encoder} &         & \multicolumn{1}{r}{9,876,576} & \multicolumn{1}{r}{985.33} \\ \hline
    \multicolumn{1}{l}{EfficientNet} & \multirow{4}{*}{LinkNet} & \multicolumn{1}{r}{32,967,082} & \multicolumn{1}{r}{1006.48} \\
    \multicolumn{1}{l}{ResNet} &                & \multicolumn{1}{r}{83,675,530} & \multicolumn{1}{r}{2852.50} \\
    \multicolumn{1}{l}{InceptionNet} &          & \multicolumn{1}{r}{31,486,746} & \multicolumn{1}{r}{672.91} \\
    \multicolumn{1}{l}{DenseNet} &              & \multicolumn{1}{r}{23,745,954} & \multicolumn{1}{r}{1491.42} \\
    \multicolumn{1}{l}{U-Net Encoder} &         & \multicolumn{1}{r}{26,924,098} & \multicolumn{1}{r}{1229.67} \\ \hline
   \multicolumn{1}{l}{EfficientNet} & \multirow{4}{*}{Feature Pyramid Network} & \multicolumn{1}{r}{9,193,322} & \multicolumn{1}{r}{810.41} \\
    \multicolumn{1}{l}{ResNet} &                & \multicolumn{1}{r}{36,592,650} & \multicolumn{1}{r}{2583.21} \\
    \multicolumn{1}{l}{InceptionNet} &          & \multicolumn{1}{r}{10,210,714} & \multicolumn{1}{r}{506.84} \\
    \multicolumn{1}{l}{DenseNet} &              & \multicolumn{1}{r}{8,026,146} & \multicolumn{1}{r}{1347.58} \\
    \multicolumn{1}{l}{U-Net Encoder} &         & \multicolumn{1}{r}{10,778,306} & \multicolumn{1}{r}{1084.12} \\ \hline
    \multicolumn{1}{l}{EfficientNet} & \multirow{4}{*}{Pyramid Attention Network} & \multicolumn{1}{r}{2,324,554} & \multicolumn{1}{r}{613.96} \\
    \multicolumn{1}{l}{ResNet} &                & \multicolumn{1}{r}{31,347,178} & \multicolumn{1}{r}{2541.75} \\
    \multicolumn{1}{l}{InceptionNet} &          & \multicolumn{1}{r}{8,696,989} & \multicolumn{1}{r}{401.89} \\
    \multicolumn{1}{l}{DenseNet} &              & \multicolumn{1}{r}{30,895,429} & \multicolumn{1}{r}{6750.59} \\
    \multicolumn{1}{l}{U-Net Encoder} &         & \multicolumn{1}{r}{3,675,426} & \multicolumn{1}{r}{812.47} \\ \hline
    \multicolumn{1}{l}{EfficientNet} & \multirow{4}{*}{U-Net Decoder} & \multicolumn{1}{r}{38,360,266} & \multicolumn{1}{r}{855.48} \\
    \multicolumn{1}{l}{ResNet} &                & \multicolumn{1}{r}{88,824,170} & \multicolumn{1}{r}{2720.53} \\
    \multicolumn{1}{l}{InceptionNet} &          & \multicolumn{1}{r}{28,645,114} & \multicolumn{1}{r}{2720.53} \\
    \multicolumn{1}{l}{DenseNet} &              & \multicolumn{1}{r}{20,213,122} & \multicolumn{1}{r}{1324.73} \\
    \multicolumn{1}{l}{U-Net Encoder} &         & \multicolumn{1}{r}{22,581,250} & \multicolumn{1}{r}{1707.23} \\ 
        \bottomrule
        \end{tabular}
%     \end{adjustbox}
%     \vspace{ - 05 mm}
\end{table*}

While deep learning demands a larger volume of data to train classification and regression networks, the collection of such data is often challenging. Data augmentation techniques, initially presented in~\cite{vandyk:2001:the_art_of_data_augmentation}, and subsequently utilised~\cite{Mokolajczyk:2018:data_augmentation,Zhang:2019:a_novel_deep_learning,Oza:2022:image_augmentation_techniques}, aim to mitigate this issue. Data augmentation is a technique for increasing the number of data samples in a dataset by modifying current samples or producing new artificial data. This tactic can assist in reducing overfitting during the training phase. In this study, we implemented several data augmentation strategies. Firstly, we adjusted the voxel values to fall between $0$ and $1$ by setting a lower threshold of $0$ and an upper threshold of $500$. Subsequently, we randomly sampled patches from the 3D volumes. Patches are randomly flipped around the x, y, or z axis with a $10\%$ probability. Similarly, they are also randomly rotated around the x, y, or z axis by 90, 180, or 270 degrees, again with a $10\%$ probability. 

\subsection{Overview of Encoders and Decoders}
We used a few complex CNNs as the encoder to enhance the performance of image segmentation models on medical images. These recent architectural models, such as EfficientNet, ResNet, InceptionNet and DenseNet, have excellent performance on classification tasks. The encoder’s duties include discovering features and delivering the initial low-resolution representations. EfficientNet is a family of models developed by Google AI, which uses a compound scaling method to efficiently scale up CNNs on layers, channels and input size~\cite{Tan:2019:EfficientNet}. ResNet uses skip connections to allow gradients to flow through the network directly, which effectively mitigates the vanishing gradient problem~\cite{He:2016:ResNet}. InceptionNet introduced Inception modules that use a mixture of filter sizes in the same layer for dimensionality reduction~\cite{Szegedy:2016:Inception}. DenseNet concatenates layer inputs and outputs, which means that the network can preserve the original information from the previous layer without being overshadowed by the output of the current layer~\cite{Huang:2017:DenseNet}.  Skip connections are used between the encoder and decoder networks or between the encoder and decoder network layers in order to prevent the vanishing gradient problem and obtain fine-grained data from the preceding layer.  

The choice of EfficientNet, ResNet, InceptionNet, and DenseNet as encoders are motivated by their proven effectiveness in feature extraction across various image recognition tasks and their widespread recognition and reliability due to extensive testing and validation in the research community. While newer architectures might offer potential advantages, their relative novelty, unestablished performance, and potentially higher computational requirements make them less preferred for this specific context. However, these newer architectures could be explored in future work to evaluate any potential improvements.  

The main goal of the image segmentation challenge is to separate the image into various segments, each representing a distinct entity. Unlike classification tasks that typically provide global image predictions, segmentation tasks involve extraction and reconstruction of feature maps to provide pixel-wise predictions, distinguishing different regions within the image. In this study, similar to the encoder part, several advanced neural networks have been applied to build the decoder part, including DeepLabV3, LinkNet, Feature Pyramid Network and Pyramid Attention Network. DeepLabV3 is an extension of DeepLabV2, which contains Atrous Convolution and up-sampled filters for dense feature extraction~\cite{chen:2017:rethinking}. LinkNet has a similar idea to ResNet that employs skip connections to directly connect layers to mitigate the gradient vanishing problem~\cite{Chaurasia:2017:LinkNet}. Feature Pyramid Network is primarily used for object detection tasks and generates a pyramid of feature maps at different scales to detect objects at various scales effectively~\cite{Lin:2017:feature_pyramid_networks}. A top-down pathway generates higher-resolution features by upscaling spatially coarser feature maps from higher layers. Pyramid Attention Network introduces a new attention mechanism based on pyramid pooling~\cite{li:2018:pyramid}. It uses both spatial and channel-wise attention in its pyramid attention module to refine the feature representation, improving performance on complex scenes. These four decoders were tested as part of the segmentation pipeline in this work. DeepLabV3, LinkNet, Feature Pyramid Network, and Pyramid Attention Net-work are chosen as decoders due to their demonstrated effectiveness in segmentation tasks and their balance between handling different scales of information. The decision was made considering performance, reliability, and computational efficiency, despite the existence of newer but less validated or more computationally demanding decoders. 

\subsection{Evaluation metrics}
In this study, we utilised two metrics, namely Dice Similarity Coefficient~\cite{Dice:1945:measures_of} and Hausdorff Distance~\cite{Gro:1915}—to assess the effectiveness of the techniques. Intuitively, segmentation performance is gauged by assessing the degree to which predictions and ground truths overlap. Results with more overlap with the ground truth are better than those with less overlap.  

\begin{equation}\label{dice_coeff}
    \text{Dice}(A,B) = \frac{2 \times |A \cap B|}{|A| + |B|}
\end{equation}

Dice Coefficient ranges from 0 to 1, where a score of 1 indicates the highest level of similarity or overlap between two point sets A and B as given in Equation~\ref{dice_coeff}. 

The Hausdorff Distance is a measure used to determine the degree of resemblance between two sets of points. It is widely used in computer vision and image processing to compare two different shapes. As shown in (2), where A and B are the two point sets compared and d(a, b) is the Euclidean distance between a and b. we used the 95th percentile Hausdorff distance (HD95), which is more robust to outliers, as commonly used in other related work. 

%d_H\{A,B\} = max {\underset{a \in A}{max}\ {\underset{b \in B}{min} \ d(x,y)
\begin{equation}
    d_H\{A,B\} = max \ \{\underset{a \in A}{max} \ \underset{b \in B}{min} \ d(x,y), \underset{b \in B}{max} \ \underset{a \in A}{min} \ d(x,y) \}
\end{equation}

%%%%%%%%%%%%%%%%%%%%%%%%%%%%%%%%%%%%%%%%%%%
%%%%%%%%%%%%%%%%%%%%%%%%%%%%%%%%%%%%%%%%%%%
\section{Experimental Results and Discussion}
The test results of the 25 segmentation models are in Table~\ref{tab:results}. More specifically, the performance of models trained using different encoders and decoders is presented using the Dice coefficient of our best experimental configurations.  

Overall, almost all the setups using the LinkNet decoder-based model produced Dice scores of $0.80$ or above, compared to roughly $0.70$ for the DeepLabV3-based model. EfficientNet-LinkNet performs the best among the configurations, with a dice of $0.8820$. Additionally, the second-placed Dice of $0.8270$ was due to ResNet and LinkNet as the encoder and decoder, respectively. For the top encoder-decoder paired model (EfficientNet-LinkNet), the HD95 is $4.7527 \pm 4.3136$. 

\begin{table*}
\centering
\caption{Dice Similarity Coefficient of various configurations. FPN: Feature Pyramid Network; PAN: Pyramid Attention Network}
\label{tab:results}
    \begin{tabular}{lccccc}
    \toprule
    \textbf{Decoders} & \multicolumn{5}{c}{\textbf{Encoders}} \\ 
    \textbf{} & \textbf{EfficientNet} & \textbf{ResNet} & \textbf{Inception} & \textbf{DenseNet} & \textbf{U-Net Encoder} \\
    \midrule
    \multicolumn{1}{l}{DeepLabV3} & \multicolumn{1}{c}{$0.796 \pm 0.049$} & \multicolumn{1}{c}{$0.782 \pm 0.059$} & \multicolumn{1}{r}{$0.788 \pm 0.042$} & \multicolumn{1}{c}{$0.794 \pm 0.044$} & \multicolumn{1}{r}{$0.796 \pm 0.045$} \\
     \multicolumn{1}{l}{LinkNet} &  \multicolumn{1}{l}{$0.882 \pm 0.013$} & \multicolumn{1}{l}{$0.827 \pm 0.064$} & \multicolumn{1}{l}{$0.807 \pm 0.050$} & \multicolumn{1}{l}{$0.803 \pm 0.053$} & \multicolumn{1}{l}{$0.782 \pm 0.047$} \\   
      \multicolumn{1}{l}{Feature Pyramid Network} &  \multicolumn{1}{l}{$0.824 \pm 0.009$} & \multicolumn{1}{l}{$0.742 \pm 0.071$} & \multicolumn{1}{l}{$0.780 \pm 0.040$} & \multicolumn{1}{l}{$0.768 \pm 0.057$} & \multicolumn{1}{l}{$0.758 \pm 0.054$} \\   
       \multicolumn{1}{l}{Pyramid Attention Network} &  \multicolumn{1}{l}{$0.776 \pm 0.050$} & \multicolumn{1}{l}{$0.756 \pm 0.057$} & \multicolumn{1}{l}{$0.764 \pm 0.049$} & \multicolumn{1}{l}{$0.727 \pm 0.065$} & \multicolumn{1}{l}{$0.702 \pm 0.047$} \\
        \multicolumn{1}{l}{U-Net Decoder} &  \multicolumn{1}{l}{$0.800 \pm 0.045$} & \multicolumn{1}{l}{$0.778 \pm 0.062$} & \multicolumn{1}{l}{$0.792 \pm 0.047$} & \multicolumn{1}{l}{$0.796 \pm 0.056$} & \multicolumn{1}{l}{$0.810 \pm 0.046$} \\   
     \bottomrule
    \end{tabular}
\end{table*}

A few reasons can explain why EfficientNet-LinkNet works better than other models. Firstly, the compound scaling technique in EfficientNet effectively scales up the network’s depth, width, and height, which can contribute to improved coronary artery segmentation. This is particularly beneficial as CTCA data contains coronary artery voxels that exist in both high and low-level features of the images. EfficientNet has different kernel sizes, which may be better at handling arteries at different resolutions. As a decoder, LinkNet is slightly better than other decoders, and this may be because LinkNet is more efficient than other decoders in up-sampling to recover the high-resolution details lost during the encoding process. The architecture of LinkNet may also enable more precise segmentation, especially of structures as delicate and detailed as coronary arteries. 

\begin{figure*}
    \centering
    \includegraphics[scale=0.7]{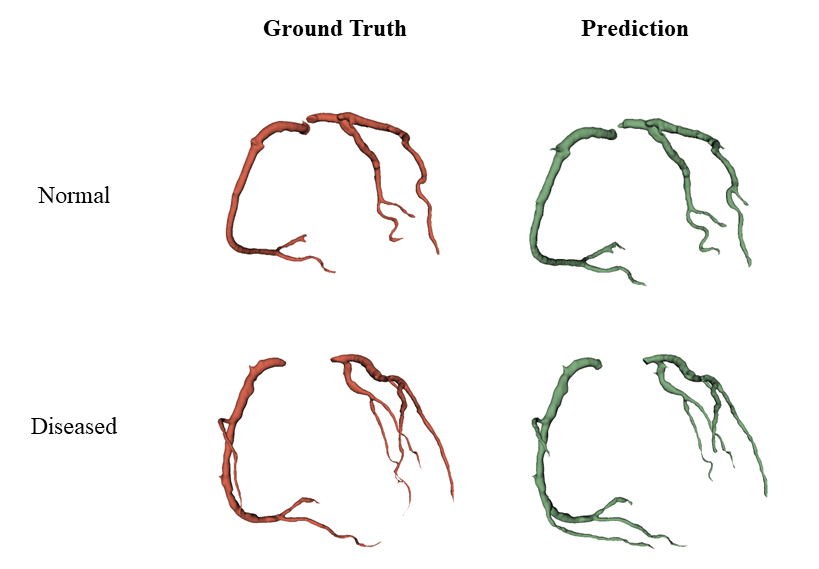}
    \caption{Segmentation results of normal versus diseased cases using the EffiicientNet-LinkNet model. Dice Similarity Coefficient (DSC) achieved are 0.890 (normal) and 0.872 (diseased).}
    \label{fig:normal_vs_diseased}
\end{figure*}

In Figure~\ref{fig:normal_vs_diseased}, we further show examples with ground truth and the prediction masks generated by EfficientNet-LinkNet. Predictions are shown in green, whereas the ground truth arteries are coloured red. Our predictions express promising segmentation performance, as seen from the results. The boundary of artery segmentation is similar to the matching ground truth, displayed in Figure~\ref{fig:normal_vs_diseased}. Overall, the arteries are segmented quite correctly. However, there are cases where additional arteries are included that are considered non-essential and have not been annotated by cardiologists. 

Our ablation study revealed findings on substituting the components of our encoder-decoder model. When we replaced the original EfficientNet encoder with a simpler U-Net structure, we observed a significant decrease in the Dice coefficient, from $0.8820$ to $0.7820$. This drastic reduction underscores the importance of EfficientNet’s efficient feature extraction capabilities, as well as its ability to capture complex patterns in the data, which are clearly critical for high-precision segmentation of coronary arteries. Moreover, replacing the decoder also led to a decrease in Dice coefficient for different architectures, including LinkNet, FPN and PAN. These observations validate the importance of both encoder and decoder parts in our network and emphasise their interdependence for achieving optimal performance. Furthermore, the results underscore the need for a powerful decoder to correctly interpret and assemble the rich feature maps produced by the encoder. Overall, the results strongly suggest that advanced architectures such as EfficientNet provide benefits for medical image segmentation tasks, and replacing such architectures with simpler ones can noticeably degrade the model’s performance. 

Compared with previous work in the MICCAI challenge~\cite{Gharleghi:2022:automated_segmentation}, our approach exhibits superior performance and a lower HD95. Unlike segmentation of abdominal organs, coronary arteries constitute much smaller regions within the CTCA volume. Consequently, coronary arteries are more likely to be overlooked during segmentation. The primary source of error in our model arises from identifying additional arteries that were not annotated by clinicians in the ground truth segmentation Figure~\ref{fig:normal_vs_diseased}. This indicates that our model can effectively detect coronary arteries. However, it also suggests that a post-processing method, such as radius-based thresholding, is necessary to eliminate unwanted structures. 

In clinical research, the quality of our segmentation is sufficient to facilitate downstream analysis, such as geometric analysis and blood flow simulation. Geometric features of arteries play a crucial role in plaque formation, which can result in severe arterial narrowing. Prior to our study, arteries were primarily manually segmented from CTCA. This approach was not only inefficient but also hampered studies aiming to uncover the true cause of disease in large populations. With our automated method, it is now possible to accurately segment arteries within a matter of minutes. Moreover, blood flow analysis is an extension of geometric analysis, where shear stresses are computed through fluid dynamics simulation. With accurate artery segmentation, it is now feasible to estimate the stresses in plaque regions, enabling the prediction of plaque progression, which may involve severe, if not fatal, plaque rupture. 

This study does have a few limitations. First, we did not explore various post-processing techniques that could potentially improve the results. Secondly, the dataset size is relatively small and the models were not tested on any other public or private datasets, which restricts our ability to assess the transferability of the model. Thirdly, we did not test transformer networks. Despite their potential as demonstrated in various domains, transformers are generally less data-efficient and often struggle to outperform CNNs, particularly with a limited dataset. We plan to address these issues in our future work. 

%%%%%%%%%%%%%%%%%%%%%%%%%%%%%%%%%%%%%%%%%%%
%%%%%%%%%%%%%%%%%%%%%%%%%%%%%%%%%%%%%%%%%%%
\section{Conclusion}
In this work, we have shown that the EfficientNet-LinkNet model exhibits good segmentation performance on the ASOCA benchmark dataset. The quality of segmentation is good enough for downstream analysis, including geometric feature extraction and blood flow simulation. The Dice of $0.882$ achieved by the best model was compared with other encoder-decoder models in an ablation study to show competitive performance. Although we improved the accuracy of coronary artery segmentation from CTCA, the problem of segmenting coronary arteries for a large population is far from being solved. In future, we plan to further improve the transferability of the method. This will involve integrating EfficientNet-LinkNet with other techniques to ensure its performance on different datasets while improving the results’ accuracy.

%%%%%%%%%%%%%%%%%%%%%%%%%%%%%%%%%%%%%%%%%%%
%%%%%%%%%%%%%%%%%%%%%%%%%%%%%%%%%%%%%%%%%%%
\section*{Acknowledgement}
This research includes computations using the computational cluster Katana supported by Research Technology Services at UNSW Sydney. This research is supported by the Australian Government Research Training Program (RTP) Scholarship, the NSW Capacity Building grant and NHMRC Ideas grant. 

%%%%%%%%%%%%%%%%%%%%%%%%%%%%%%%%%%%%%%%%%%%
%%%%%%%%%%%%%%%%%%%%%%%%%%%%%%%%%%%%%%%%%%%
\section*{Declaration of competing interest}
The authors declare no conflicts of interest.

%%%%%%%%%%%%%%%%%%%%%%%%%%%%%%%%%%%%%%%%%%%
%%%%%%%%%%%%%%%%%%%%%%%%%%%%%%%%%%%%%%%%%%%
\bibliography{references.bib}
\bibliographystyle{IEEEtran}

%%%%%%%%%%%%%%%%%%%%%%%%%%%%%%%%%%%%%%%%%%%
%%%%%%%%%%%%%%%%%%%%%%%%%%%%%%%%%%%%%%%%%%%
\end{document}